\documentclass[aps,prl,twocolumn,showpacs,preprintnumbers,amsmath,amssymb]{revtex4}

\usepackage{graphicx}
\usepackage{amsmath}
\usepackage{amsfonts}
\usepackage{color}

\begin{document}
\title{Thermodynamic density of states of two-dimensional GaAs systems
near the apparent metal-insulator transition}

\author{G.D.Allison, E.A.Galaktionov, A.K.Savchenko, and S.S.Safonov}
\affiliation{School of Physics, University of Exeter, Exeter EX4 4QL, UK}
\author{M.M.Fogler}
\affiliation{Department of Physics, University of California San Diego, La
Jolla, California 92093,USA}
\author{M.Y.Simmons}
\altaffiliation[Present address: ]{University of New South Wales,
Sydney, Australia 2052.}
\author{D.A.Ritchie}
\affiliation{Cavendish Laboratory, University of Cambridge, Madingley Road,
Cambridge CB3 0HE, UK}

\begin{abstract}

We perform combined resistivity and compressibility studies of
two-dimensional hole and electron systems which show the apparent
metal-insulator transition -- a crossover in the sign of $\partial
R/\partial T$ with changing density. No thermodynamic anomalies have been
detected in the crossover region. Instead, despite a ten-fold difference in
$r_s$, the compressibility of both electrons and holes is well described by
the theory of nonlinear screening of the random potential. We show that the
resistivity exhibits a scaling behavior near the percolation threshold
found from analysis of the compressibility. Notably, the percolation
transition occurs at a much lower density than the crossover.

\end{abstract}

\maketitle

%%%%%%%%%%%%%%%%%%%%%%%%%%%%%%%%%%%%%%%%%%%%%%%%%%%%%%%%%%%%%%%%%%%%

The apparent metal-insulator transition (MIT) in high-mobility
two-dimensional systems remains a topic of fundamental
interest~\cite{Abrahams_01} and continuing
debate~\cite{Finkelstein_05_DasSarma_05}. The anomaly of these systems is
exemplified by the existence of a narrow range of carrier densities around
$n=n_c$ where the slope of the temperature dependence of the resistance,
$\partial R /\partial T$, changes its sign. To unravel a complex interplay
between interactions and disorder in this phenomenon, it is essential to
combine transport measurements with other experimental probes, in
particular measurements of the thermodynamic density of states (also
referred to as the charge compressibility ~\cite{Eisenstein_94,Shapira_96})
$\chi = d n / d\mu$, where $\mu$ is the chemical potential. There have been
only few measurements of $\chi$ near the apparent MIT
~\cite{Dultz_00,Ilani_00, Kravchenko_03}, among which work~\cite{Dultz_00}
on a 2D hole gas with large values of the Coulomb interaction parameter
$r_s \equiv 1 / \sqrt{\pi n a_B^2} \approx 5 - 16$ has attracted much
attention. (Here $a_B = 18\,{\rm \AA}$ is the effective Bohr radius for the
hole mass of $0.38~m_0$.) In their experiments done at $T=0.3-1.3$~K the
authors of Ref.~\cite{Dultz_00} found that the inverse compressibility $\chi^{-1}(n)$ has a minimum
which is positioned exactly at $n_c$. This was interpreted as a
thermodynamic signature of an interaction-driven phase transition discussed
in theoretical works~\cite{Si_98,Chakravarty_99}.

An alternative explanation of the minimum of $\chi^{-1}(n)$ can be based on the
nonlinear screening theory
(NST)~\cite{Efros_Shklovskii_book,Efros_88_93,Shi_02,Fogler_04} that
emphasizes the role of disorder. The basic premise of the NST is that a
low-density metal is unable to screen fluctuations of potential, so that
depletion regions with vanishingly small local density appear and grow as
$n$ decreases. The NST predicts that $\chi^{-1}(n)$ has a minimum at $n =
n_m$ (determined by disorder), and a rapid upturn to positive
values at $n < n_m$.

This theory also predicts a percolation threshold at $n = n_p$~\cite{Efros_88_93}, where $n_p \approx n_m / 3$ in typical GaAs systems~\cite{Fogler_04}. There have been suggestions, based on the
conductance scaling, that the percolation transition is closely related to
the change in the sign of $\partial R /\partial T $
\cite{Meir_99,Shi_02,DasSarma_05}. (The existence of the percolative MIT in
2D GaAs structures was proposed earlier in \cite{Efros_88_93}.)

In this work we use combined compressibility and conductance measurements
to shed light on the origin of the apparent MIT in 2D hole gases with large
interactions between the carriers -- a problem widely debated over the last
few years \cite{Proskuryakov_01_02, Noh_03, Leturcq_03}. We compare the
behaviour of the compressibility of holes with that of electrons (with much
weaker interactions) and conclude that the apparent MIT in 2D hole gases is
\emph{not} related to a quantum phase transition. Similar to the case of 2D
electrons, it is most likely caused by an interplay of different scattering
mechanisms~\cite{Lilly_DasSarma_03, Gold_86, Zala_01}. The
combined measurements also enable us to demonstrate that there is \emph{no}
direct link between the MIT and the percolation transition either.

Two types of system have been examined: a 2DEG with $r_s =1-4$ and a 2DHG
with $r_s = 10-35$. We find \emph{no} relation between the position $n_m$
of the minimum in $\chi^{-1}$ and the position $n_c$ of the MIT -- the two
densities can differ by a factor of two and, furthermore, their ratio is
sample and cooldown dependent. The $\chi^{-1}(n)$ dependence fits the
NST~\cite{Fogler_04} predictions very well for \emph{both} carrier types.
We show that in some range of $T$ the dependence of the conductance of both
systems on carrier density fits the usual percolation scaling
ansatz. The found percolation threshold agrees with the
prediction from the analysis of the compressibility in terms of the NST.
However, its density is significantly lower than the crossover density
$n_c$, which makes a direct, universal connection between the apparent MIT (the sign change of $\partial R /\partial T$) and percolation doubtful.

\begin{figure}
\includegraphics[width=0.4\textwidth]{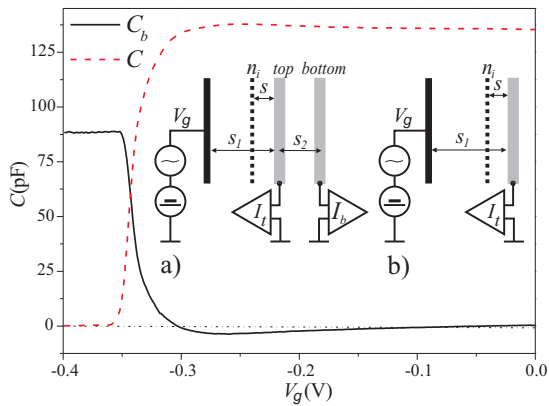}
\caption{Capacitance of a double-layer 2DEG structure E01 measured
simultaneously by two methods: $C_b$ -- field-penetration, $C$ -- capacitance;
$T=4.2$ K, $f=4$ Hz. Inset: diagrams of capacitance measurements in (a) double-
and (b) single-layer structures. \label{nexper} }
\end{figure}

%%%%%%%%%%%%%%%%%%%%%%%%%%%%%%%%%%%%%%%%%%%%%%%%%%%%%%%%%%%%%%%%%%%%%%%%%%%
Our 2DEG structures E01 and E02 contain two GaAs quantum wells of width
$w=200$~\AA~separated by a $200\,{\rm \AA}$ and $300\,{\rm \AA}$-thick AlGaAs
barrier, respectively, Fig.~\ref{nexper}. The top-layer mobility at the highest density is
$5\times10^5$ cm$^2/$Vs for E01 and $8\times10^5$ cm$^2/$Vs for E02. The
2DHG samples H03 and H05 with mobility $4.7\times10^5$ cm$^2/$Vs and
$5.6\times10^5$ cm$^2/$Vs, respectively, are formed in standard
single-layer AlGaAs/GaAs heterostructures with Au gates.

The resistance as a function of gate voltage, $V_g$, is measured at
$T=0.03 - 10\,{\rm K}$, in a similar way to ~\cite{Proskuryakov_01_02},
and the relation between $V_g$ and $n$ is established by the Hall effect.
The position of $n_c$ is determined at the lowest temperature of the
compressibility measurements, $T = 0.26$~K. The compressibility is found
to be temperature independent in the range $0.26-1.5$~K for 2DHGs and
$0.26-5$~K for 2DEGs. The compressibility and conductance measurements have
been repeated several times, one after another, to confirm that there is no
drift in the structure. The compressibility is determined using two
techniques -- the ``capacitance'' and the ``field-penetration'',
Fig.~\ref{nexper}.

{\it Capacitance method\/} --- Here the ac voltage $\tilde{V_g} = 2\,{\rm
mV}$ of frequency $f = 1 - 100$~Hz is applied to the gate and the
$90^{\circ}$ phase-shifted current $\tilde{I}$ in the probed layer is
measured (Fig.~\ref{nexper}, inset), yielding the capacitance $C =
\tilde{I} / 2\pi f\tilde{V_g}$. Instead of $\chi$ we discuss the quantity
which is a measure of inverse compressibility: the Thomas-Fermi screening
radius $d \equiv (\varepsilon\varepsilon_0 / e^2)
\chi^{-1}$~\cite{Eisenstein_94}. It is easy to show that $d$ is related to
$C$ by the formula $ d = \varepsilon\varepsilon_0 A (C^{-1} - C_0^{-1}) /
(1 + \gamma^{-1}) $, where $A$ is the gated area and $C_0 = \varepsilon
\varepsilon_0 A / s_1$ is the geometric capacitance between the gate and
the probed 2D layer, Fig~\ref{nexper}. In a double-layer structure, where the ``probed'' 2D
layer is the top quantum well, the factor $\gamma \equiv (d_b + s_2 + w) /
s_1$ accounts for the electrostatic interaction between the two
layers~\cite{Eisenstein_94}. Here $s_2$ is the separation between the
centers of the quantum wells and $d_b \approx a_B/4 \ll s_2 + w$ is the
screening radius of the bottom layer. In a single-layer structure
$s_2\rightarrow \infty$ and the correction $\gamma^{-1}$ is set to zero.

{\it Field-penetration method\/} --- In this technique the compressibility
of the top layer is found from the capacitance $C_b$ between the gate and
the bottom layer. The measured capacitive current $\tilde{I}_b$ is caused
by the electric field penetrating through the top layer in
proportion to its $\chi^{-1}$. Once $C_b(V_g)$ is found, the screening
radius $d$ is calculated using Eqs.~(7)--(9) of Ref.~\cite{Eisenstein_94}.

In real 2D systems there is a change in the transverse confinement (and the
corresponding subband energy) with varying $n$. This gives a small
contribution $\Delta d_{\mathrm{sub}}$ to $d$ whose sign and magnitude depend on the
structure and the method of capacitance measurement. To compare the results
obtained in different experimental situations, we subtract this
contribution and discuss the quantity $d^* = d - \Delta d_{sub}$. The value
of $\Delta d_{\mathrm{sub}}$ is of the order of the 2D layer thickness and is
calculated by a perturbation theory. For a single-layer heterostructure
$\Delta d_{\mathrm{sub}} = 0.46 (a_B / n)^{1/3}$ \cite{Ando_82}. For double-layer
(quantum well) structures we compute $\Delta d_{\mathrm{sub}}$ using the infinite
square-well approximation, similarly to Ref.~\cite{Combescot_93}. This
gives for the capacitance method $\Delta d_{\mathrm{sub}} = +0.3967 w (1 - 0.0544
\lambda) $ in the thin-well limit, $\lambda = n w^3 / a_B \ll 1$. In the
field-penetration technique $\Delta d_{\mathrm{sub}} = -0.1033 w (1 - 0.1148
\lambda) $, cf.~Ref.\cite{Eisenstein_94}.

\begin{figure}
\includegraphics[width=0.4\textwidth]{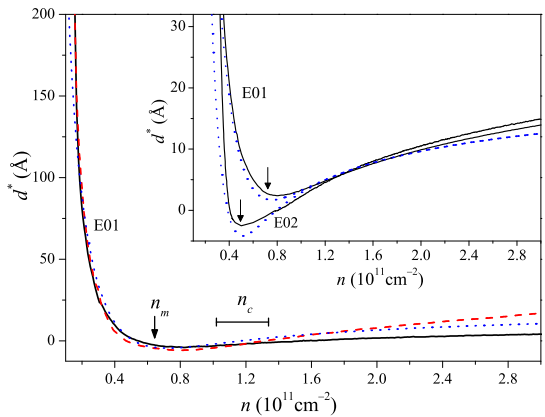}
\caption{$d^{*}$ for 2DEG structure E01 with $s=200$~\AA, $s_1=700$~\AA, $s_2=400$~\AA. Solid lines are the results of the field-penetration method, dashed line - the capacitance method. Dotted line - the NST theory with $n_i=1.2\times10^{11}$~cm$^{-2}$. (The donor concentration found from $\tau_q(n)$ is $n_i^\tau=2.3-4.5\times10^{11}$~cm$^{-2}$, while the maximum concentration of (uncorrelated) donors is $n_i^G=9.0\times10^{11}$~cm$^{-2}$.) Inset: Comparison with 2DEG structure E02 ($s=400$\AA, $s_1=900$~\AA, $s_2=500$~\AA), where $n_i=2.2\times10^{11}$~cm$^{-2}$ and $n_i^\tau=3-4.5\times10^{11}$~cm$^{-2}$, while $n_i^G=11\times10^{11}$~cm$^{-2}$). \label{ndt}}
\end{figure}

In Fig.~\ref{ndt} we present $d^*$ for sample E01 found from both the
capacitance and the field-penetration methods. The field-penetration technique
is more accurate than the capacitance technique at large $n$, because in this
method the subtraction of the geometric term $C_0$ is obviated. However, at low
densities around the minimum in $\chi^{-1}$ the contribution $C_0$ is less
important, and one can see that the two methods give nearly identical results.

Now we turn to the comparison of the results with the nonlinear screening
theory. This theory predicts:
\begin{equation}
d^* = (a_B/4) + \Delta d_{\mathrm{ex}} + \Delta d_{\mathrm{cor}} + \Delta d_{\mathrm{dis}}.
\label{d_full}
\end{equation}
The first term in Eq.~(\ref{d_full}) is due to the single-particle density
of states (kinetic energy) of the 2D carriers. The correction $\Delta d_{\mathrm{ex}} = -(8 \pi^3 n)^{-1/2}$ comes from the exchange interaction. Another
negative contribution is due to correlations between the carriers: $\Delta d_{\mathrm{cor}} = (\varepsilon \varepsilon_0 / e^2) \mathrm{d}^2 (n E_c)/ \mathrm{d} n^2$, where the
correlation energy per particle $E_c$ is computed according to
Ref.~\cite{Tanatar_89}. As the carrier density decreases, $\Delta d_{\mathrm{ex}}$
and $\Delta d_{\mathrm{cor}}$ cause a change in the sign of $d$ from positive to
negative -- the effect seen
experimentally~\cite{Eisenstein_94,Dultz_00,Kravchenko_03}. Disorder,
however, brings a positive contribution, $\Delta d_{\mathrm{dis}}$, responsible for
the upturn of $\chi^{-1}$ at low densities~\cite{Fogler_04}:
\begin{equation}
\Delta d_{\mathrm{dis}} = \frac{3\sqrt{2}}{32\pi^2}
\frac{(0.3+\eta) s}{0.036 \eta +0.12 \eta^3 + \eta^3}
\exp(-4\pi\eta^2),
\label{EQDR}
\end{equation}
where $\eta \equiv n s / \sqrt{n_i}$, $s$ is the spacer and $n_i$ is the
effective 2D concentration of dopants (an adjustable parameter, see below).

Equation~(\ref{EQDR}) was derived assuming that disorder is produced by a
$\delta$-doped layer of \emph{uncorrelated\/} dopants with a
two-dimensional concentration $n_{i}$~\cite{Fogler_04}. The samples in this
work have three-dimensional doping. It can be shown that the impurities
closest to the 2D layer have the greatest effect on $\chi$, and therefore
Eq.~(\ref{EQDR}) is still valid provided one uses an effective $n_i$.
Unfortunately, $n_i$ cannot be determined in a simple way because not all
impurities can be ionized and also because of existing correlations in
their positions ~\cite{Buks_94}. Both factors reduce the effective
$n_i$~\cite{Efros_88_93} compared to a naive estimate based on the total
number of impurities known from the growth conditions (denoted here by
$n_i^G$). A better estimate of $n_i$ is deduced from the quantum lifetime
$\tau_q$. We have found $\tau_q(n)$ from the analysis of the Shubnikov-de
Haas effect and determined the effective concentration $n_i^\tau$ according
to Ref.~\cite{Gold_86} -- it turns out to be 3-5 times smaller than $n_i^G$.

The dotted line in Fig.~\ref{ndt} is the best fit by Eq.~(\ref{d_full}) for
structure E01, with $n_i$ found to be close to $n_i^\tau$. Good agreement
with Eq.~(\ref{d_full}) where $n_i$ is close to $n_i^\tau$ has also been
obtained for E02 -- the inset compares the results for the two 2DEG
structures. The latter has a larger spacer and in agreement with the NST
the $\chi^{-1}(n)$ minimum, $n_m\approx0.38~n_i^{1/2}/s$, is shifted to
lower $n$. The position of the minimum in $\chi^{-1}(n)$, indicated by an arrow, is found from the
fitted curve with an accuracy of $\lesssim2\%$. (Inaccuracy of
the calculation of $\Delta d_{\mathrm{sub}}$ has an even smaller effect on the position
of $n_m$.) The range of densities where $\partial R /\partial T$ changes its sign is indicated in
Fig.~\ref{ndt} by $n_c$. It is seen that the crossover in $R(T)$ occurs at higher
densities than the minimum in $\chi^{-1}(n)$.

\begin{figure}
\includegraphics[width=0.4\textwidth]{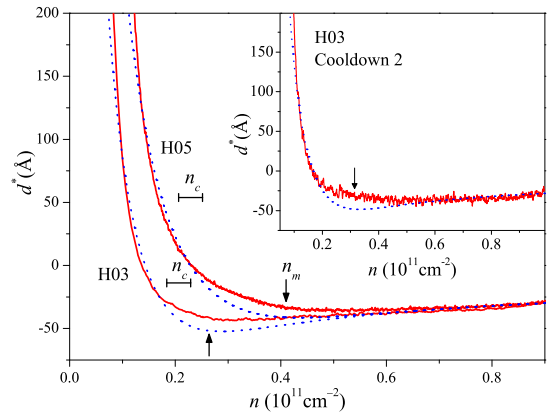}
\caption{$d^{*}$ for 2DHG structures H03 and H05 with $s=500$~\AA~and $s_1=2670$~\AA, obtained by the capacitance method. Dotted lines - NST theory. The dopant concentrations for H03 are: $n_i=1\times10^{11}$~cm$^{-2}$, $n_i^\tau=1.1-2.2\times10^{11}$~cm$^{-2}$ and $n_i^G=5.6\times10^{11}$~cm$^{-2}$. The minimum for H05 ($n_i=2.4\times10^{11}$~cm$^{-2}$, $n_i^G=5.6\times10^{11}$~cm$^{-2}$) is shifted to larger $n$. Inset: $d^{*}$ for H03 (cooldown 2), $n_i=1.5\times10^{11}$~cm$^{-2}$ and $n_i^G=5.6\times10^{11}$~cm$^{-2}$. \label{pdt} }
\end{figure}

Figure~\ref{pdt} shows the results for 2DHG samples H03 (two cooldowns)
and H05. The dotted lines are best fits to Eq.~(\ref{d_full}) with
parameters $n_i$ consistent with our analysis of $\tau_q(n)$. In
the second cooldown the upturn of $d^*$ occurs at higher densities and the
obtained $n_i$ is also larger, i.e., the sample in this cooldown is more
disordered. This supports the notion of correlations among the impurities
since these are known to depend on thermal cycling~\cite{Buks_94}. As in
the 2DEG structures, the apparent MIT region $n_c$ does not overlap with
$n_m$, although here $n_c < n_m$. This is not surprising because of the
difference in the contributions to $R(T)$ of electrons and holes that determine the position of
$n_c$. The same conclusion that $n_c < n_m$ was drawn for another studied
2DHG sample (H06, not shown).

%%%%%%%%%%%%%%%%%%%%%%%%%%%%%%%%%%%%%%%%%%%%%%%%%%%%%%%%%%%%%%%%%%%%%%%%%%%
%%%%%%%%%%%%%%%%%%%%%%%%%%%%%%%%%%%%%%%%%%%%%%%%%%%%%%%%%%%%%%%%%%%%%%%%%%

\begin{figure}
\centerline{
\includegraphics[width=0.47\textwidth]{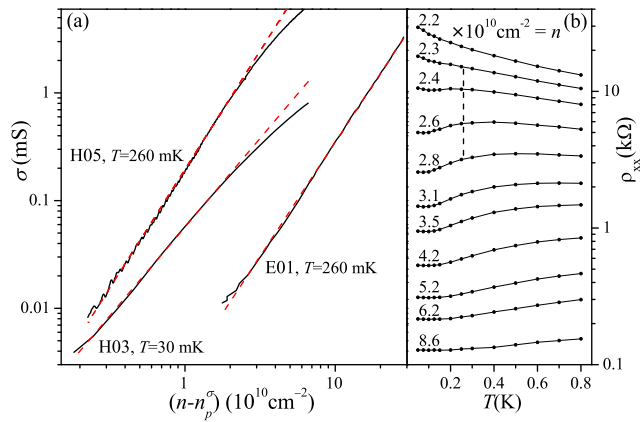}
} \caption{ (a) Fits of the conductance to $\sigma(n) = \alpha (n -
n_p^\sigma)^t$. (b) An example of the temperature dependence of the
resistivity of H03 structure. The range of $n_c$ in this cooldown is
indicated by the dotted line. \label{fig:percolation_fit} }
\end{figure}

To understand the relation between the apparent MIT and percolation in
2DHGs, we have attempted to extract the percolation threshold $n_p^\sigma$
from the fit of the conductance to $\sigma(n) = \alpha (n - n_p^\sigma)^t$
following the procedure in \cite{Meir_99,Leturcq_03,DasSarma_05}.
Figure~\ref{fig:percolation_fit} shows the results
with the exponent $t = 2.1\pm0.1$ at $T=0.26$~K for both electrons and
holes, which is close to that in works \cite{Leturcq_03,DasSarma_05}. For the 2DHGs the
exponent decreases with decreasing temperature to
$t = 1.6\pm0.1$, which is close to $t = 1.31$ expected for classical
percolation~\cite{Grassberger_99}. Notably, the value of the percolation
threshold is significantly lower than $n_c$: $n_c / n_p^{\sigma} = 2.9$ for
H03 and $n_c / n_p^{\sigma} = 2.1$ for H05. Therefore, we surmise that the
apparent MIT is not due to percolation but, similar to
2DEGs~\cite{Lilly_DasSarma_03}, is due to an interplay of the ``metallic''
$T$-dependence caused by phonon scattering and Fermi-liquid corrections to
impurity scattering, and the ``insulating'' dependence caused by
localization~\cite{Proskuryakov_01_02, Noh_03, Gold_86, Zala_01}.

We want to stress, however, that using the percolation scaling in a quantum
system requires justification. One can rationalize this procedure by
arguing that the percolation concept can apply in some intermediate range
of $T$, high enough to preempt localization at $n > n_p^\sigma$ yet low
enough to inhibit thermal activation at $n < n_p^\sigma$. Another point of
concern is that the scaling is observed over a broad range of densities (up
to $n/n_p^\sigma\sim~2$, similar to \cite{Meir_99,DasSarma_05}), while it
is expected to work only near the critical point \cite{Gold_91}. However,
our combined transport and compressibility measurements allow us to examine
directly the applicability of the scaling procedure and demonstrate its
validity. Using the NST prediction~\cite{Efros_88_93,Fogler_04}, we obtain
independently $n_p \approx 0.12 \sqrt{n_i} / s \approx n_m / 3$ and compare
it with $n_p^\sigma$. We have established that the percolation thresholds
derived from both methods are very close: $n_p = 0.76\times
10^{10}\,{\rm cm}^{-2}$ and $n_p^{\sigma} = 0.7\times 10^{10}\,{\rm
cm}^{-2}$ for H03, and $n_p = 1.2 \times 10^{10}\,{\rm cm}^{-2}$ and
$n_p^{\sigma} = 1 \times 10^{10}\,{\rm cm}^{-2}$ for H05. (This comparison
is done at $T=0.26$~K and even better agreement is obtained at $T=30$~mK as
$n_p^{\sigma}$ increases with decreasing $T$ by about 20\%.)

%%%%%%%%%%%%%%%%%%%%%%%%%%%%%%%%%%%%%%%%%%%%%%%%%%%%%%%%%%%%%%%%%%%%%%%%%%%
%%%%%%%%%%%%%%%%%%%%%%%%%%%%%%%%%%%%%%%%%%%%%%%%%%%%%%%%%%%%%%%%%%%%%%%%%%

In summary, our combined conductance and compressibility measurements
suggest that the apparent MIT in 2DHGs with $r_s$ up to $\sim 30$ is
neither an interaction-driven phase transition nor a percolation
transition. The behavior of the compressibility at low hole densities is
well described by the nonlinear screening theory. This indicates that the upturn in
$\chi^{-1}(n)$ is due to depletion regions in the channel, with total area less than $3.5$\%, caused by
disorder.

Support from the ORS Award (E.A.G.) and the Hellman and the Sloan Foundations
(M.M.F.) is gratefully acknowledged. We thank S.~Das Sarma, H.~W.~Jiang, M.~P. Lilly and
Y.~Meir for useful discussions.

%%%%%%%%%%%%%%%%%%%%%%%%%%%%%%%%%%%%%%%%%%%%%%%%%%%%%%%%%%%%%%%%%%%%%%


\begin{thebibliography}{10}

\bibitem{Abrahams_01}
E.~Abrahams {\it et~al.},
\newblock Rev. Mod. Phys. {\bf 73}, 251 (2001);
S.~V. Kravchenko and M.~P. Sarachik,
\newblock Rep. Prog. Phys. {\bf 67}, 1 (2004).

\bibitem{Finkelstein_05_DasSarma_05}
A.~Punnoose and A. Finkel'stein,
\newblock Science {\bf 310}, 289 (2005);
S.~{Das Sarma} and E.~H. Hwang,
\newblock Solid. State Commun. {\bf 135}, 579 (2005).

\bibitem{Eisenstein_94}
J.~P. Eisenstein {\it et~al.},
\newblock Phys. Rev. B {\bf 50}, 1760 (1994).

\bibitem{Shapira_96}
S.~Shapira {\it et~al.},
\newblock Phys. Rev. Lett. {\bf 77}, 3181 (1996).

\bibitem{Dultz_00}
S.~C. Dultz and H.~W. Jiang,
\newblock Phys. Rev. Lett. {\bf 84}, 4689 (2000).

\bibitem{Ilani_00}
S.~Ilani {\it et~al.},
\newblock Phys. Rev. Lett. {\bf 84}, 3133 (2000).

\bibitem{Kravchenko_03}
M.~Rahimi {\it et~al.},
\newblock Phys. Rev. B {\bf 67}, 081302 (2003).

\bibitem{Si_98}
Q.~Si and C.~M. Varma,
\newblock Phys. Rev. Lett. {\bf 81}, 4951 (1998).

\bibitem{Chakravarty_99}
S.~Chakravarty {\it et~al.},
\newblock Phil. Mag. B {\bf 79}, 859 (1999).

\bibitem{Efros_Shklovskii_book}
A.~L. Efros and B.~I. Shklovskii,
\newblock {\em Electronic Properties of Doped Semiconductors},
\newblock (Springer, New York, 1984).

\bibitem{Efros_88_93}
A.~L. Efros,
\newblock Solid. State Commun. {\bf 70}, 253 (1988);
%\emph{ibid.} {\bf 70}, 253 (1988).
%
%\bibitem{Efros_93}
A.~L. Efros, F.~G.~Pikus, and V.~G.~Burnett,
\newblock Phys. Rev. B {\bf 47}, 2233 (1993).

\bibitem{Shi_02}
J.~Shi and X.~C. Xie,
\newblock Phys. Rev. Lett. {\bf 88}, 086401 (2002).

\bibitem{Fogler_04}
M.~M. Fogler,
\newblock Phys. Rev. B {\bf 69}, 121409(R) (2004).

\bibitem{Meir_99}
Y.~Meir,
\newblock Phys. Rev. Lett. {\bf 83}, 3506 (1999) and experiments cited therein.

\bibitem{DasSarma_05}
S.~{Das Sarma} {\it et~al.},
\newblock Phys. Rev. Lett. {\bf 94}, 136401 (2005).

\bibitem{Proskuryakov_01_02}
Y.~Y.~Proskuryakov {\it et~al.},
\newblock Phys. Rev. Lett. {\bf 86}, 4895 (2001); \emph{ibid.}
{\bf 89}, 076406 (2002).

\bibitem{Noh_03}
H.~Noh {\it et~al.},
\newblock Phys. Rev. B {\bf 68}, 165308 (2003).

\bibitem{Leturcq_03}
R.~Leturcq {\it et~al.},
\newblock Phys. Rev. Lett. {\bf 90}, 076402 (2003).

\bibitem{Lilly_DasSarma_03}
M.~P. Lilly {\it et~al.},
\newblock Phys. Rev. Lett. {\bf 90}, 056806 (2003).

\bibitem{Gold_86}
A.~Gold and V.~T. Dolgopolov,
\newblock Phys. Rev. B {\bf 33}, 1076 (1986);
S.~Das~Sarma and E.~H.~Hwang,
\newblock Phys. Rev. Lett. {\bf 83}, 164 (1999).

\bibitem{Zala_01}
G.~Zala, B.~N. Narozhny, and I.~L. Aleiner,
\newblock Phys. Rev. B 64, 214204 (2001).

\bibitem{Ando_82}
T.~Ando, A.~B.~Fowler, and F~Stern,
\newblock Rev. Mod. Phys. {\bf 54}, 437 (1982).

\bibitem{Combescot_93}
M.~Combescot {\it et~al.},
\newblock Solid State Commun. {\bf 88}, 309 (1993).

\bibitem{Tanatar_89}
B.~Tanatar and D.~M. Ceperley,
\newblock Phys. Rev. B {\bf 39}, 5005 (1989).

\bibitem{Buks_94}
E.~Buks {\it et~al.},
\newblock Phys. Rev. B {\bf 49}, 14790 (1994);
%
%\bibitem{Koenraad_97}
P.~M. Koenraad {\it et~al.},
\newblock Superlatt. and Microstr. {\bf 21}, 237 (1997);
%
%\bibitem{Zhitinev_00}
N.~B. Zhitenev {\it et~al.},
\newblock Nature (London) {\bf 404}, 473 (2000).

\bibitem{Grassberger_99}
P.~Grassberger,
\newblock Physica A {\bf 262}, 251 (1999).

\bibitem{Gold_91}
A.~Gold,
\newblock Phys. Rev. B {\bf 44}, 8818 (1991).
\end{thebibliography}
\end{document}